\documentclass[fleqn,usenatbib]{mnras}

\usepackage{newtxtext,newtxmath}
\usepackage{siunitx}

\usepackage[T1]{fontenc}

\DeclareRobustCommand{\VAN}[3]{#2}
\let\VANthebibliography\thebibliography
\def\thebibliography{\DeclareRobustCommand{\VAN}[3]{##3}\VANthebibliography}

\usepackage{graphicx}	
\usepackage{amsmath}	
\usepackage{booktabs}
\usepackage{adjustbox}
\usepackage{subcaption}
\usepackage{float}
\usepackage{makecell}
\usepackage{longtable}
\usepackage{adjustbox}

\newcommand{\mgii}{Mg~{\sc ii}~}
\newcommand{\feii}{Fe~{\sc ii}~}

\newcommand{\kms}{$km s^{-1}$}

\newcommand{\OII}{\mbox{[O\,{\sc ii}]}}

\newcommand{\MgI}{\mbox{Mg\,{\sc i}}}
\newcommand{\MgII}{\mbox{Mg\,{\sc ii}}}

\newcommand{\FeII}{\mbox{Fe\,{\sc ii}}}

\newcommand{\Magiicat}{\textsc{MAGIICAT}}

\title[CGM of quasar host galaxies]{Circumgalactic medium of quasar host galaxies at $0.4\le z\le 0.8$ probed by strong \MgII\ absorption}

\author[]{
Paryag Sharma,$^{1}$\thanks{panditparyag@gmail.com}
Raghunathan Srianand,$^{2}$
Hum Chand,$^{1}$
Labanya Kumar Guha$^{3}$
\\
$^{1}$Department of Physics and Astronomical Science, Central University of Himachal Pradesh, Dharamshala, 176215, India\\
$^{2}$Inter-University Centre for Astronomy and Astrophysics, Post Bag 4, Ganeshkhind, Pune 411 007, India\\
$^{3}$Indian Institute of Astrophysics, II Block, Koramangala, Bengaluru-560 034, India
}

\date{Accepted XXX. Received YYY; in original form ZZZ}

\pubyear{\the\year{}}

\begin{document}
\label{firstpage}
\pagerange{\pageref{firstpage}--\pageref{lastpage}}
\maketitle

\begin{abstract}
Using a sample of 166 projected quasar pairs, we investigate the influence of active galactic nuclei on the circumgalactic medium (CGM) of the quasar host galaxies probed using strong \MgII\ absorption (i.e., \(W_{2796}\ge 1\)\AA) at impact parameters ($D$) $<$100 kpc. The foreground quasars are restricted to the redshift range \(0.4 \leq z \leq 0.8\) and have median bolometric luminosity and stellar mass of $10^{45.1} erg~s^{-1}$and $10^{10.89} M_\odot$ respectively.  We report detections of \MgII\ absorption in 29 cases towards the background quasar and in 4 cases along the line of sight to the foreground quasars.
%
We do not find any difference in the distribution of \(W_{2796}\) and covering fraction (\(f_c\)) as a function of $D$ between quasar host galaxies and control sample of normal galaxies. 
These results are different from what has been reported in the literature, possibly because: (i) our sample is restricted to a narrow redshift range, (ii) comparative analysis is carried out after matching the galaxy parameters, (iii) we focus mainly on strong \MgII\ absorption and (iv) our sample lacks foreground quasars with high bolometric luminosity (i.e., $L_{bol}>10^{45.5}$ erg s$^{-1}$). Future studies probing luminous foreground quasars, preferably at lower impact parameters and higher equivalent width sensitivity is needed to  consolidate our findings.

%
%
\end{abstract}

\begin{keywords}
quasars: absorption lines -- quasars: general -- galaxies: haloes
\end{keywords}



\section{Introduction}
\label{sec:introduction}
Since quasars were first discovered, their spectra revealed metal absorption lines produced by the intervening gas \citep{Sandage1965ApJ...141.1560S,Gunn1965ApJ...142.1633G,Burbidge1966ApJ...144..447B,Burbidge1967ARA&A...5..399B}. Subsequent studies established that these metal absorbers were associated with foreground galaxies intersecting the quasar sightlines at low impact parameters \citep{Bahcall1966ApJ...144..847B,Bahcall1969ApJ...156L..63B,Opher1974Natur.250..310O,Roeser1975A&A....45..329R,Burbidge1977ApJ...218...33B,Bahcall1978PhyS...17..229B,Sargent1982ApJ...256..374S,Bergeron1986A&A...169....1B,Steidel1992ApJS...80....1S}. This realization has led to the identification of an extended gaseous component surrounding galaxies. 
Various statistical properties of Mg~{\sc ii} absorbers can be reproduced using this gaseous component, referred to as  
the circumgalactic medium (CGM), around galaxies \citep[e.g., see,][]{Petitjean1990,Srianand1994}.
The CGM also plays a crucial role in galaxy evolution, acting as a reservoir for gas that fuels star formation and accretion processes. 
Understanding the CGM’s properties and its interaction with galactic components is essential for constructing a comprehensive picture of galaxy dynamics \citep{Tumlinson2017ARA&A..55..389T,Fumagalli2024arXiv240900174F}.

Detecting diffused emission-lines allows direct spatial mapping of the gas and, in principle, provides an alternate probe of the CGM \citep[e.g., see,][]{Cantalupo2014Natur.506...63C,Martin2014ApJ...786..106M,Burchett2021ApJ...909..151B,Das2024MNRAS.52710358D}. However, such studies face significant challenges compared to absorption-line studies as the emission measure scales with the square of the gas density (\(n^2\)), since the CGM typically has a low hydrogen number density (\(n_H \sim 10^{-2} \, \text{cm}^{-3}\) or less), detecting such emission is inherently difficult. Additionally, the surface brightness of CGM emission is extremely faint relative to the sky and detector backgrounds, and surface brightness dimming increases steeply with redshift, further complicating detection. Nevertheless, this limitation is substantially alleviated in the presence of bright quasars, where enhanced ionizing radiation can significantly boost the surface brightness of both resonant and non-resonant line transitions. Recent deep observations have revealed that giant Ly$\alpha$ nebulae are common around luminous quasars at $z \sim 2$–$5$, demonstrating that the quasar radiation field can illuminate the cool CGM to detectable levels \citep[e.g., see,][]{Borisova2016ApJ...831...39B,Arrigoni_Battaia2018MNRAS.473.3907A,Fossati2021MNRAS.503.3044F}.

Observations have shown that the Ly$\alpha$ surface brightness of circum-quasar nebulae correlates with quasar luminosity, indicating that photoionization by the central AGN plays a major role in powering the emission \citep{Mackenzie2021MNRAS.502..494M}.
At lower redshifts (i.e., $z\sim0.4$–$1.4$), \OII-emitting circumgalactic nebulae have been detected around UV-luminous quasars, further supporting the presence of dense, metal-enriched gas reservoirs \citep{Johnson2024ApJ...966..218J,Dutta2023MNRAS.522..535D}. In addition, detailed mapping has shown that the morphologies and kinematics of these nebulae arise from a range of processes, including galaxy interactions \citep{Helton2021MNRAS.505.5497H,Liu2024MNRAS.527.5429L}, cool gas accretion \citep{Johnson2022ApJ...940L..40J}, radio jet interaction \citep[e.g.,][]{Shukla2022} and feedback-driven outflows exhibiting elevated velocity dispersions \citep{Liu2025arXiv250312597Z}.
These emission studies provide crucial constraints on the structure, kinematics, and physical conditions of the CGM in quasar host halos.

In contrast, quasar absorption-line studies offer several advantages. 
The absorption-line detections are largely independent of redshift and the luminosity of the host galaxy and depends mainly on the signal to noise  ratio (SNR) achieved for the quasar spectrum. 
Therefore, by choosing bright enough background sources high sensitivity can be achieved to detect even low column density gas.
On the flip side, absorption-line studies provide only projected, pencil-beam measurements of gas surface density. In most cases, observations are limited to a single sightline per galaxy due to the rarity of suitably bright background quasars, making it difficult to fully characterize the spatial extent and morphology of the CGM. Therefore, one generally uses statistics based on a sample of quasar-galaxy pairs to characterize the average properties of CGM around galaxies.

Quasar absorption-line studies have yielded profound insights into the nature of CGM \citep[e.g.,][]{Nielsen2013ApJ...776..114N,Mishra2018MNRAS.473.5154M,Dutta2020MNRAS.499.5022D,Dutta2021MNRAS.508.4573D,Dutta2023MNRAS.522..535D,Joshi2024arXiv241207835J}.
Early efforts to use quasar pairs to probe the CGM of quasars itself were led by \citet{Bowen2006ApJ...645L.105B}, who demonstrated that strong absorption lines such as Mg\,\textsc{ii} can be detected from the CGM of quasars. These studies were expanded by the Quasars Probing Quasars (QPQ) program \citep[e.g.,][]{Hennawi2006ApJ...651...61H,Prochaska2013ApJ...776..136P,Lau2016ApJS..226...25L} and references therein, which focused on $z \gtrsim 2$ quasars and revealed an overdensity around the quasars 
\citep[see, also][]{Jalan2019ApJ...884..151J,Jalan2021MNRAS.505..689J}.
More recently, studies at lower redshifts, such as \citet{Farina2014MNRAS.441..886F} and \citet{Johnson2015MNRAS.452.2553J}, showed that quasars at $z \sim 1$ have enhanced Mg\,\textsc{ii} absorption compared to normal galaxies, and that the CGM covering fraction correlates with quasar luminosity.
Building on these results, \citet{Chen2023ApJS..265...46C} performed a systematic stacking based analysis of Mg\,\textsc{ii} absorption around quasars in SDSS DR16Q \citep{Lyke}, finding a strong decline in absorber incidence with impact parameter and evidence for anisotropy relative to the quasar radiation field.\par
In this study we adopt the absorption-line technique to probe the CGM because the advent of large-scale spectroscopic surveys like the Sloan Digital Sky Survey (SDSS; \citealt{sdssdr172022ApJS..259...35A}) and the Dark Energy Spectroscopic Instrument (DESI; \citealt{DESI2024AJ....168...58D}) has dramatically increased the number of known quasars, providing a significant sample in a narrow redshift range suitable for statistical analysis. In particular, we ask the question whether the properties of CGM, as probed by \MgII\ absorption, are same for normal galaxies and host galaxies of quasars.
%
Since the samples used in previous studies span a wide range of redshifts and luminosities, comparisons to normal galaxies is not straight forward due to degeneracies inherent in such samples (e.g luminosity dependence vs. z-dependence if most of the high luminosity objects come from high redshifts). To enable a proper comparison between normal galaxies and galaxies hosting a quasar, a controlled investigation is needed, which serves as the main motivation for our work.

This paper is organized as follows: Section~\ref{sec:sample_selection} describes our sample and absorption line measurements. Section~\ref{sec:results} presents our analysis and results, while Section~\ref{sec:discussion} we present discussion and conclusion of our findings.
Throughout this work, we assume a flat \(\Lambda\)CDM cosmology with \(\Omega_m = 0.3\), \(\Omega_\Lambda = 0.7\), and \(h_0 = 0.7\).

\section{Sample and absorption line measurements}
\label{sec:sample_selection}
The sample of projected quasar pairs used in this study is sourced from the Sixteenth Data Release of the Sloan Digital Sky Survey \citep[DR16Q;][]{Lyke}, comprising 750,414 quasars. Given the signal-to-noise limitations of large survey spectra, we concentrate our analysis on strong intervening \ion{Mg}{II} absorbers. We restrict the impact parameter to be less than 100 kpc to focus on the inner regions of the CGM, where strong \mgii absorption is most likely to occur. Beyond this scale, \mgii absorption tends to be significantly weaker with much lower covering factor.
Moreover, this limit help us to ensure that the detected absorption is most likely to be arising from a single galaxy halo, minimizing contamination from neighboring galaxies or large-scale clustering, though such effects are unlikely in view of the recent under density observed around quasars \citep{Shibata2025arXiv250504259S}. To minimize the effects of redshift evolution in our study, we focus on quasar pairs where the foreground quasar falls within the redshift range \(0.4 \leq z \leq 0.8\). The lower redshift limit of \(z = 0.4\) ensures that \ion{Mg}{II} absorption remains within the optical spectral range, while this redshift interval also corresponds to a regime where the equivalent width versus impact parameter distribution is well sampled for normal galaxies. Therefore, we apply the following criteria to construct our sample:

\begin{enumerate}
    \item  We first demand that the transverse separation between quasar sight lines at the redshift of the foreground quasar is less than 100 kpc
    and is satisfied only by 1115 projected quasar pairs.
    

    \item We restricted the redshift of the foreground quasar to the range $0.4 \leq z_f \leq 0.8$, reducing the sample to 202 projected pairs.

    \item A minimum velocity offset of 6000 \kms\ is imposed between the redshifts of the foreground and background quasars to reduce the likelihood of quasar pairs being associated  \citep[as also used in previous study by][]{Chen2023ApJS..265...46C}. This further reduces the size of our sample to 180 projected quasar pairs.
    
    \item To avoid contamination of the \mgii absorption  by high-z Ly$\alpha$ forest in the background quasar spectra, we imposed the condition $(1 + z_f) \times 2800 \geq (1 + z_b) \times 1300$, where $z_f$ and $z_b$ are the redshifts of the foreground and background quasars respectively. After applying this conditions we are left with 149 projected quasar pairs.
\end{enumerate}
One pair, featuring the foreground quasar J083649.46+484150.1 from \cite{Bowen2006ApJ...645L.105B}, was also included in our sample.
Additionally, we performed the same sample selection exercise on the DESI EDR \citep{DESI2024AJ....168...58D} and identified a total of 16 projected quasar pairs
satisfying all the above constraints. We also searched for projected quasar pairs where one quasar is from the DESI sample and the other from the SDSS sample, but did not find any pairs that satisfied our selection criteria. Thus our final sample consists of 166 quasar pairs.

\begin{figure}
    \centering
    \includegraphics[width=\linewidth]{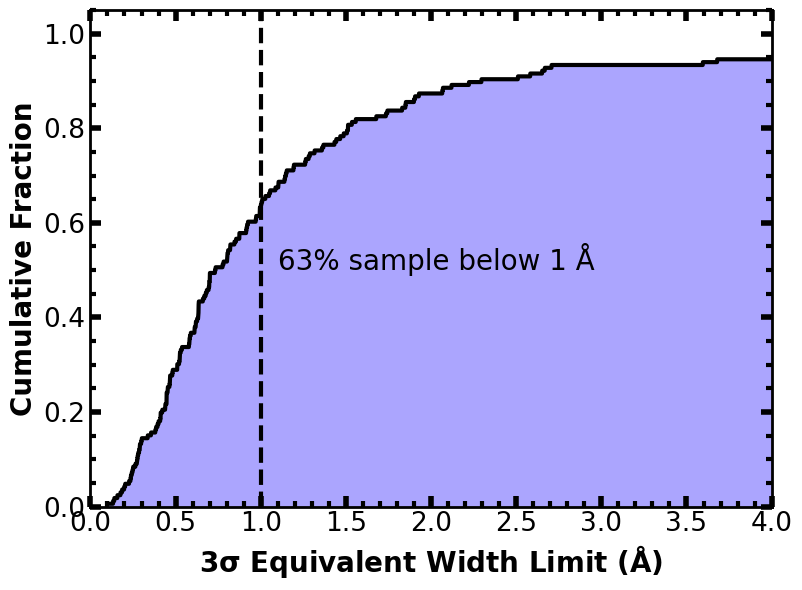}
    \caption{Cumulative distribution of the \(3\sigma\) limit of $W_{2796}$ in the spectrum of background quasars at the location of the foreground quasars. Vertical dashed line marks the rest equivalent limit of 1\AA.
    }
    \label{fig:ew_limits_distribution}
\end{figure}

To start with, we measure the 3$\sigma$ upper limit of \mgii\ rest equivalent width ($W_{2796}$) in the spectrum of the background quasar within $\pm$1500 \kms\ with respect to the foreground quasar's redshift, following the procedure described by \citet{Churchill2000ApJ...543..577C}. For this purpose, we have used 
line free regions after clipping the absorption lines, if present, in the above mentioned velocity range. Cumulative distribution of the $3\sigma$ limiting \MgII\ rest equivalent width for the full sample is shown in Figure~\ref{fig:ew_limits_distribution}. It is evident from the figure that only 63\% (i.e.,  105 sightlines) of the background quasar spectra have sufficient SNR  so that we can detect
\MgII\  absorption with $W_{2796}\ge1$\AA\ at $>3 \sigma$ level. For the statistical analysis of direct detections we use these projected quasar pairs and a limiting threshold $W_{2796}$ of 1\AA.

\begin{figure}
    \centering
    \includegraphics[width=\linewidth]{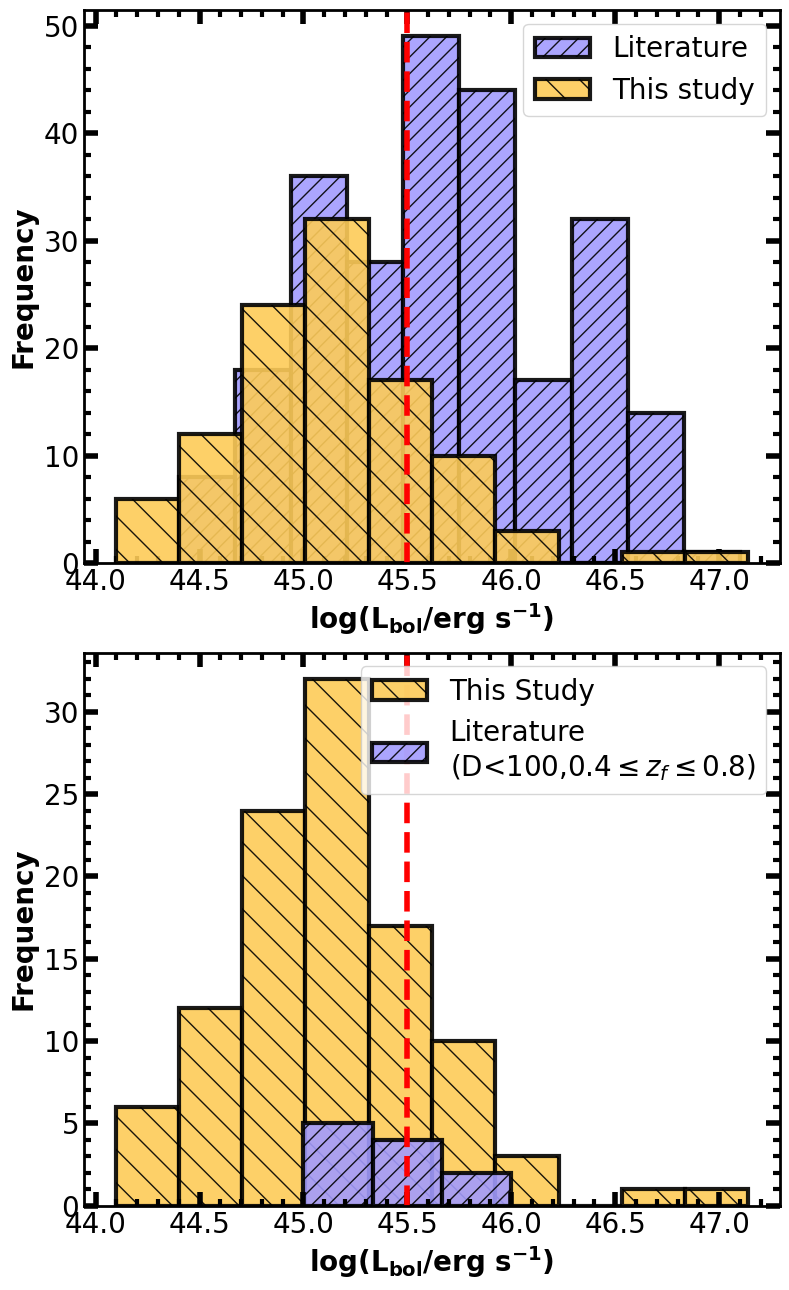}
\caption{Comparison of the bolometric luminosity of quasars in our sample with the full compiled sample from \citet{Johnson2015MNRAS.452.2553J}, which includes data from \citet{Bowen2006ApJ...645L.105B}, \citet{Farina2014MNRAS.441..886F}, \citet{Prochaska2014ApJ...796..140P}. We label this combined dataset as "Literature" in the figure legend. {\it Top Panel}: Shows the comparison with their full sample. {\it Bottom Panel}: Focuses on quasars within the same redshift and impact parameter ranges considered in our study. The vertical dashed line marks $log(L_{bol}/ergs^{-1}) = 45.5$, the division between the luminous and low luminosity quasars used in the literature sample.}

    \label{fig:luminosity_comparison}
\end{figure}

Properties of \MgII\ absorption are known to depend on the bolometric luminosity ($L_{bol}$) of the foreground quasars \citep[e.g.,][]{Johnson2015MNRAS.452.2553J}. 
 We have used the bolometric luminosities for SDSS quasar pair from \citet{Wu2022ApJS..263...42W}, and also used their method for  estimation of $L_{bol}$ of our the DESI quasars as well.
In the top panel of Figure~\ref{fig:luminosity_comparison} we compare the bolometric luminosity of foreground quasars in our sample with those from \citet{Johnson2015MNRAS.452.2553J}. It is clearly evident that compared to their sample, quasars in our sample typically probe the low bolometric luminosity range. In particular while studying the luminosity dependent effects \citet{Johnson2015MNRAS.452.2553J} defined the low luminosity objects as the ones with $\log L_{bol}(erg~s^{-1})<45.5$ (marked with vertical dashed line in the Figure \ref{fig:luminosity_comparison}). As can be seen from the figure only   20 per cent of our quasars have $L_{bol}$ in excess of $10^{45.5}~erg~s^{-1}$. 
This is an important factor to remember while comparing our results with that of \citet{Johnson2015MNRAS.452.2553J}.
In the bottom panel of Figure~\ref{fig:luminosity_comparison} we compare the distribution of $L_{bol}$ between the two samples if we restrict ourselves to $D\le 100$ kpc and foreground redshift range from 0.4 to 0.8.
Clearly our sample provides a substantial improvement at low redshift and low bolometric luminosity range compare to the sample of \citet{Johnson2015MNRAS.452.2553J}.

We identify \MgII\ doublet features in the background quasar spectrum within $\pm$1500~\kms~of the foreground quasar redshift \citep[as per the criteria in][]{Chen2023ApJS..265...46C} by visual inspection. 
These were confirmed when the \MgII\ doublet ratio is in the allowed range and by the presence of additional associated absorption from species such as \MgI\ and \FeII.
%
%
We detected \mgii\ doublets along 29 background quasar lines of sight within $\pm~1500$~\kms~ from $z_f$. Out of these 29 detections, 18 detections have their measured \MgII\ rest equivalent widths significant at \(>3\sigma\) level and in the remaining cases the measured equivalent widths are at \(\sim2\sigma - 3\sigma\) level.
%
In Table \ref{tab:catalog} we provide the details of the projected quasar pairs,  impact parameter (D) at the redshift of the foreground quasar, \mgii\ rest equivalent width or its 3$\sigma$ upper limit in case of non-detections and absorption redshift. 
For completeness, we also examined quasar pairs with a velocity separation of less than 6000 \kms. Among 22 such pairs, we detected \mgii absorption in 3 cases. However, these pairs are excluded from our analysis, as the absorption could originate either from the CGM of the foreground quasar or from the associated absorption of the background quasar.  
We also searched for \ion{Mg}{II} absorption in the spectra of foreground quasars within a 3000 \kms\ window blue-ward of their redshift to identify any line-of-sight absorption. \MgII\ absorption is detected in only four out of 166 cases.


Publicly available systemic redshifts of SDSS quasars are known to exhibit systematic biases of $\Delta z / (1 + z) \geq 0.002$ (i.e., $\geq 600$ \kms). 
\citet{Hewett2010MNRAS.405.2302H} have improved the accuracy by a factor $\sim$ 20 using Ca~{\sc ii} absorption from the host galaxy and low ionization emission lines. In the SDSS DR16 catalog, revised redshifts are provided for quasars where systematic corrections were applied; however, in some cases, only visually inspected redshifts are reported, which are less reliable. To ensure consistency and accuracy, we adopted the improved systematic redshifts for foreground quasars provided in  \citet{Wu2022ApJS..263...42W}.
To obtain accurate absorption redshifts, we utilized \texttt{VoigtFit} \citep{Krogager2018arXiv180301187K} to simultaneously fit the \mgii\ 2796, 2803 and \feii\ 2600 absorption lines. In cases where the signal-to-noise ratio (SNR) was low and simultaneous fitting was not feasible, we fitted only the \mgii\ lines, selecting one or both depending on the line profiles.  
 To compare the properties of \MgII\ absorption from the quasar host galaxies in our sample and normal galaxies at $0.4\le z \le 0.8$, we use the data from \citet[][refer to as \Magiicat\ sample]{Nielsen2013ApJ...776..114N}  and \citet{Huang2021MNRAS.502.4743H} for normal galaxies and refer to them collectively as "galaxy sample". We note that this galaxy sample comprises a mix of both star-forming and quiescent galaxies. However, due to significant uncertainties in classifying the nature of quasar host galaxies, we do not distinguish between star-forming and quiescent types in our analysis.

\begin{figure}
    \centering
    \includegraphics[width=\linewidth]{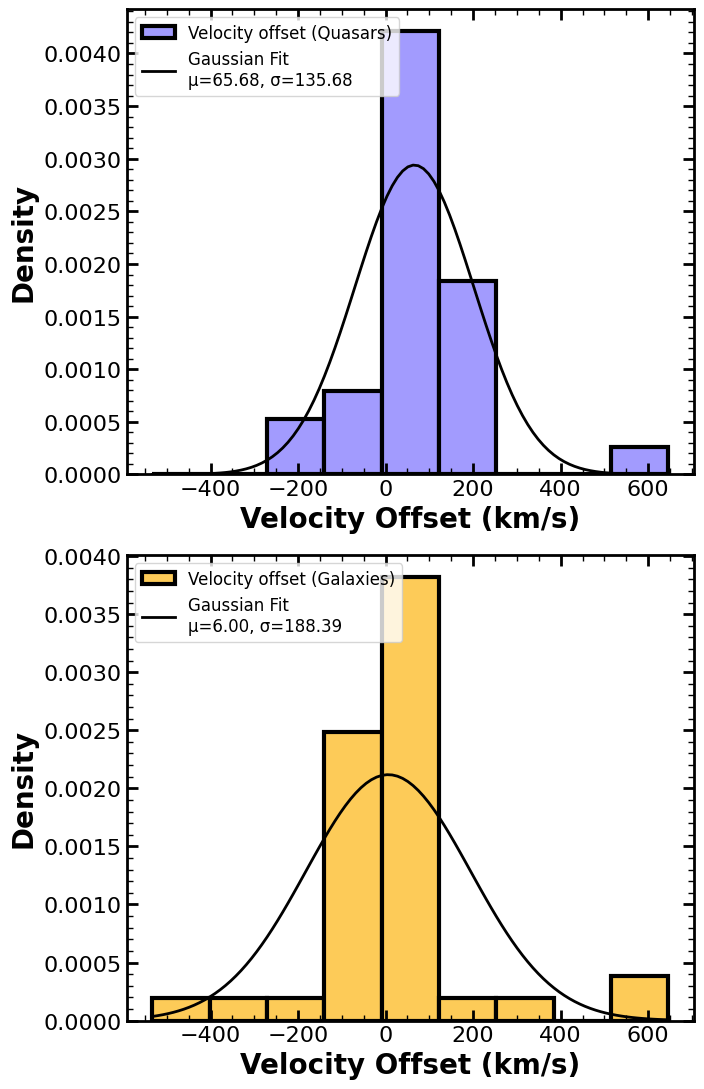}
    \caption{{\it Top panel}: Displays the velocity offset between the redshift of the foreground quasar and the \mgii absorption redshift, highlighting a strong correlation between the two, except mean offset quasar reading. {\it Bottom panel}: Illustrates the same velocity offset for galaxies with $D < 100$ and $0.4 \le z \le 0.8$, providing further context for the observed trends.}
    \label{fig:velocity_offset}
\end{figure}

In Figure~\ref{fig:velocity_offset}, the top panel illustrates the velocity offsets of the detected absorbers relative to the foreground quasar. The bottom panel of Figure~\ref{fig:velocity_offset} presents the velocity offset distribution for the galaxy sample within the selected redshift and impact parameter ranges. 
We performed a KS test between distribution of velocity offset between two samples. The initial analysis yielded a KS statistic of 0.44 with a corresponding $p_{null}$ of 0.002, suggesting a statistically significant difference between the two distributions.
However, it is important to consider the potential impact of the quasar's inherent velocity offset, calculated as $\mu = 65.68$ \kms, which may originate from systematic uncertainties in obtaining systemic redshift of the foreground quasars. After accounting for this offset, a subsequent KS test returned a statistic of 0.186 with a $p_{null}$ of 0.528. This reduces the statistical dissimilarity to a significant extent. We note that the r.m.s. velocity dispersion of the quasar sample is 135 km~s$^{-1}$, which is smaller than the 188 km~s$^{-1}$ measured for the galaxy sample. The latter is also substantially higher than the 84 km~s$^{-1}$ reported for a galaxy sample by \citet{Huang2021MNRAS.502.4743H}. This discrepancy could be attributed to differences in sample selection. In \citet{Huang2021MNRAS.502.4743H}, a "galaxy-first" approach was employed, where galaxies were identified prior to searching for associated \ion{Mg}{II} absorption. This method typically yields a lower r.m.s. velocity dispersion. 
In contrast, our galaxy sample is also based on the MAGIICAT dataset, which has less precise galaxy redshifts and incorporates a mixture of both "galaxy-first" and "absorption-first" selection strategies. In the latter case, galaxies are identified after detecting absorption features, which increases the likelihood of associating absorbers with bright galaxies in the field. This may introduce a systematic velocity offset. 
Notably, our quasar sample is conceptually similar to the "galaxy-first" approach (aside from redshift uncertainties), which could also help explain the relatively lower r.m.s. velocity dispersion compared to our galaxy sample.

\begin{table*}
    \centering
    \begin{tabular}{|lllllllll|}
    \hline
    No. & \makecell{Background \\ Quasar} & \makecell{Redshift \\ $z_b$} & \makecell{Foreground \\ Quasar} & \makecell{Redshift \\ $z_f$} & \makecell{$D_\perp$ \\ (kpc) at $z_f$} & $z_{abs}$ & \makecell{$W_{2796}$ (\AA) \\ \mgii\ 2796}&\makecell{$Err~W_{2796}$ (\AA) \\ \mgii\ 2796} \\ 
    \hline
    1 & J003715.09+055029.2 & 2.096 & J003715.29+055022.7 & 0.517 & 44.155 & 0.517 & 0.72 & 0.35\\ 
    2 & J012906.68+004322.9 & 1.838 & J012906.33+004322.0 & 0.743 & 38.928 & 0.742 & 2.43 & 1.03\\ 
    3 & J013416.95+330854.5 & 1.136 & J013416.69+330855.9 & 0.706 & 25.514 & --    & 0.39  & -1  \\ 
    \vdots & \vdots & \vdots & \vdots & \vdots & \vdots & \vdots & \vdots & \vdots \\ 
    \hline
    \end{tabular}
    \caption{Table displaying the information about our quasar pair sample. The cases in which there is a detection of \mgii\ absorption line at the foreground redshift have the equivalent width specified. For cases where the line has not been detected, we provide the $3\sigma$ $W_{2796}$ threshold and the error associated with is -1.  The entire table consisting of 166 quasar pairs is available online; here we display its structure and contents.}
    \label{tab:catalog}
\end{table*}


\section{Results}
\label{sec:results}

In this section we summarize our results of comparing properties of \mgii absorption originating from quasar host galaxies and normal galaxies at $0.4\le z\le 0.8$ and at impact parameters $D\le100$ kpc.

\subsection{Equivalent width vs. Impact parameter}
\label{sec:Equivalent width vs Impact parameter}
\begin{table}
    \centering
    \caption{Comparison of $\alpha$, $\beta$ and $\sigma$ values for \mgii $W_{2796}$ vs. D relation (e.g., see Figure~\ref{fig:Ewvskpcsep}).} 
    \begin{tabular}{cccc}
        \hline
        \makecell{Sample\\($0.4 \le  z\le 0.8)$} & $\alpha$ & $\beta$ & $\sigma$ \\ 
        \hline
        \hline
     \makecell{Quasar} & $-0.010  \pm 0.001$ & $0.295 \pm 0.039$ & $0.346 \pm 0.034$ \\
        \makecell{Galaxies} & $-0.011 \pm 0.002$  & $0.315 \pm 0.052$ & $0.691 \pm 0.075$ \\
        \hline
    \end{tabular}
    \label{tab:alpha_beta_sigma}
\end{table}

\begin{figure}
    \centering
    \includegraphics[width=\linewidth]{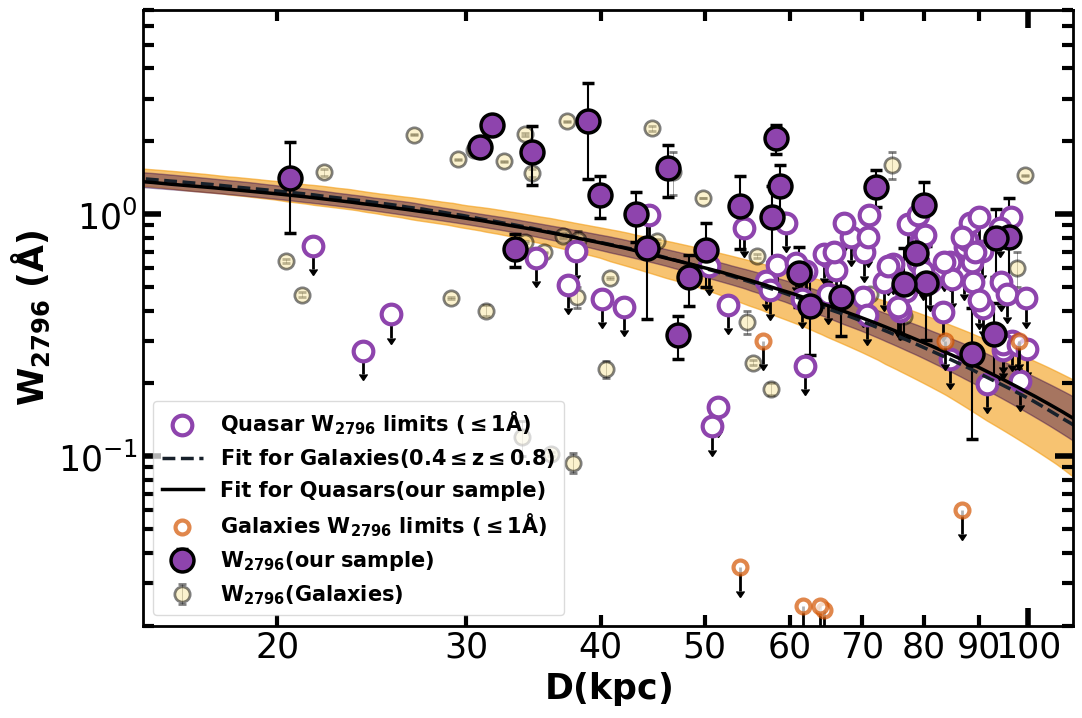}
    \caption{\( W_{2796} \) vs. impact parameter (D) for foreground quasars. Violet filled circles indicate detections; open circles with arrows show upper limits. Galaxy detections are shown as faded yellow circles; upper limits as open orange circles. The solid black line with violet shading shows the quasar fit; the dashed black line with yellow shading shows the galaxy fit over the same redshift range.}

    \label{fig:Ewvskpcsep}
\end{figure}

The rest equivalent width of \MgII\ absorption is known to show clear anti-correlation with the impact parameter \citep{Bergeron1991A&A...243..344B} albeit with a large scatter \citep{Chen2010ApJ...714.1521C,Nielsen2013ApJ...776..114N}.  This relationship is usually approximated by the log-linear relationship as,
\begin{equation}
    \log W_{2796} (\text{\AA}) = \alpha \times D\ (\text{kpc}) + \beta
\end{equation}
where $\beta$ is $\log W_{2796}$(\AA)  at $D$ = 0 {kpc} and $\alpha$ provides the exponential scale length. 
We performed the fit among the quasar and galaxy samples following the standard procedure described in \citet{Guha2022}\footnote{For non-detections, the likelihood integral is computed with a lower bound of $W_{2796} = 0$~\AA.} by using the detections as well as upper limits from non-detections, to compare their best-fit parameters: the slope ($\alpha$), intercept ($\beta$), and intrinsic scatter ($\sigma$). The best fit is shown  in Figure~\ref{fig:Ewvskpcsep} and  best-fit parameter values  are summarized in Table~\ref{tab:alpha_beta_sigma}. 
As can be seen in Table~\ref{tab:alpha_beta_sigma}, the best fit values  of $\alpha$ and $\beta$ for the quasar and galaxy samples are consistent within error-bar. However, the best fit value of $\sigma$ among the two sample differ significantly.
This difference here can be  primarily driven by the range of equivalent widths probed in each sample: the galaxy sample includes both strong and weak absorbers down to $W_{2796} = 0.3$~\AA, whereas the quasar sample includes strong absorbers with detection upper limits greater than or equal to 1\AA. 
Therefore, the difference in value of $\sigma$ obtained for two samples can be attributed to the effect of instrumental sensitivity rather than any differences in the underlying CGM properties. 
To further quantify this dependence, we performed a test using the full galaxy sample, omitting redshift and impact parameter cuts in order to improve upon the statistics by utilizing a larger sample size.
To simulate the effect of reduced sensitivity in the galaxy sample—comparable to that of the quasar sample with a minimum detected $W_{2796}$ of $\sim 0.3$~\AA—we modified the galaxy sample by replacing all detections with $W_{2796} < 0.3$~\AA, along with the upper limits, with randomly selected upper limits drawn from the quasar sample. Additionally, we excluded one high value outlier ($W_{2796} > 3$~\AA). The best-fit scatter in this sample is found to be $0.454 \pm 0.018$, showing a deviation of approximately $6\sigma$ compared to its original (higher sensitivity sample) value of $0.740 \pm 0.043$.
The best-fit values of the slope ($\alpha$) and intercept ($\beta$) exhibit only marginal changes, with their values of $-0.011 \pm 0.001$ and $0.261 \pm 0.010$, respectively. These remain consistent with the original values of $-0.011 \pm 0.002$ and $0.315 \pm 0.052$ within the $1\sigma$ uncertainty range.
These results show  that sensitivity limits contribute significantly to the observed scatter and that reducing the sensitivity brings the galaxy sample scatter from $0.740 \pm 0.043$ to $0.454 \pm 0.018$ which is closer to the scatter in quasar sample ($0.346 \pm 0.034$). However, a residual difference remains, suggesting additional contributing factors. A forward-modeling approach \citep[e.g.,][]{Huang2021MNRAS.502.4743H} may better account for these effects and will be explored in future work.

As discussed in \citet{Guha2024} the relationship between $W_{2796}$ and impact parameter (D) is sensitive to the distribution of impact parameter probed. For example, inclusion of Ultra-Strong \MgII\ absorbers (USMgII) and Galaxy On Top of Quasars (GOTOQs) that probe typical impact parameters $\le$~30 kpc makes $\beta$ values increase by a factor of 1.7 compared to the literature data that usually probe larger impact parameters \citep[see][]{Guha2022,Guha2023MNRAS.519.3319G,Guha2024,Guha2024MNRAS.532.3056G}.  Note both our quasar and galaxy samples used here have very few sightlines at $D<20$ kpc. This could explain why our $\beta$ values are less than the best fit values obtained by \citet{Guha2024}.

\begin{figure}
    \centering
    \includegraphics[width=\linewidth]{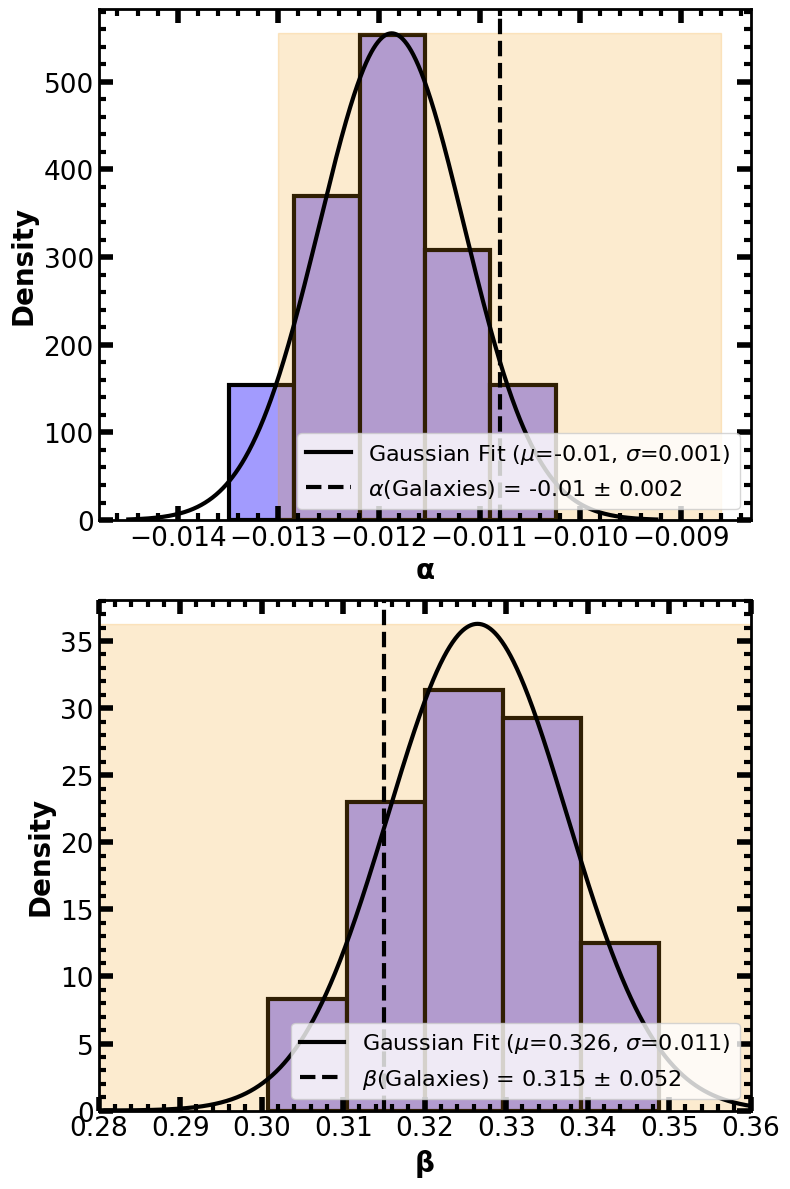}
    \caption{Figure shows the distributions of $\alpha$ ({\it top panel}) and $\beta$({\it bottom panel}) for various realizations of control sample of quasars matching in impact parameter of galaxy data with impact parameter tolerance of 10 kpc. The results indicate that both distributions are consistent with each other, suggesting no significant discrepancies in the parameter values despite the imposed selection criteria. The vertical dashed lines represent the $\alpha$ and $\beta$ values for the galaxy sample, while the shaded region indicates the associated error.
    }
    \label{fig:alpha_beta_dist}
\end{figure}

Further, we note that the distribution of the impact parameter $D$ in our quasar sample differs significantly from that of the galaxy sample ($p_{null}$ = 0), primarily because the quasar sample includes a larger number of data points at higher impact parameters. To investigate whether this discrepancy biases our results, we resampled the quasar data based on the impact parameter distribution of the galaxy sample, ensuring an equal number of points in each subset. This resampling was performed multiple times. In Figure~\ref{fig:alpha_beta_dist}, we plot the distributions of $\alpha$ and $\beta$ obtained from these resampled quasar datasets and compare them with the corresponding fit for the galaxy sample (including upper limits) within our redshift range. The medians of the Gaussian fits are found to be approximately equal to the values of $\alpha$ and $\beta$ obtained for the galaxy sample within measurement uncertainties. This indicates that the similarity observed in the $W_{2796}$ vs. $D$ relation between the galaxy and quasar samples is not biased by the differences in their impact parameter distributions. 

It was also suggested that at a given impact parameter the measured $W_{2796}$ might depend on the galaxy luminosity \citep[see discussions presented in,][]{Guha2024}. Therefore, next we explore whether such a trend is present with respect to the bolometric luminosity of quasars,
We also searched for a possible correlation between $log(W_{2796}) -\alpha \times log(D)$ and $L_{bol}$. We did not find any significant correlation between the two quantities (Spearman Rank correlation:$-0.037$, $p_{null}$: $0.704$). This implies that the scatter in $W_{2796}$ vs. $D$ plot is not driven by the dependence of this relationship on $L_{bol}$, at least in the sample considered here. 
It is worth noting that \citet{Chen2023ApJS..265...46C}, using stacked spectra, reported a dependence of rest equivalent width on luminosity. However, their higher-luminosity sample has, $\log L_{\text{bol}}/\text{erg}~\text{s}^{-1} \ge 45.66$, with most of the bright quasars coming from high-$z$ range.
Lack of such bright quasars in our sample prevents us from any direct comparison with their study.

%

\subsection{\mgii covering fraction}
\label{sec:covering fraction}
\begin{figure}
    \centering
     \includegraphics[width=1.05\linewidth]{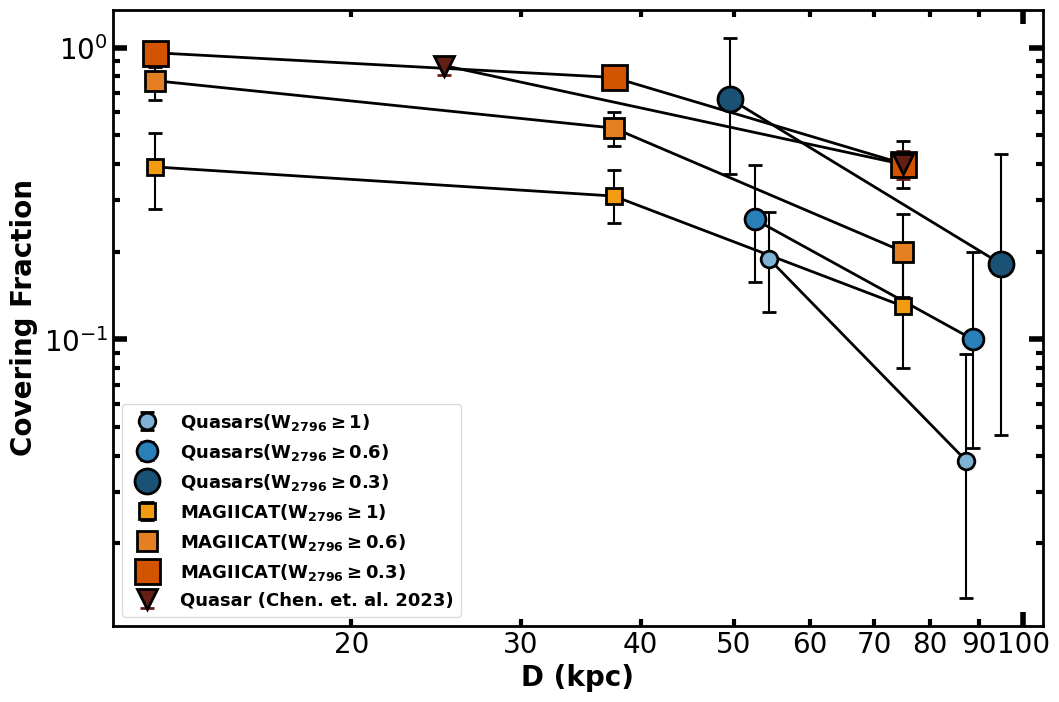}
    \caption{Comparison for the covering fraction of the  CGM of galaxies hosting quasar in  our sample with that of (i) similar study (i.e., quasar hosting galaxies)  by \citet{Chen2023ApJS..265...46C} and (ii) the  CGM study of normal galaxies in MAGICAT by \citet{Nielsen2013ApJ...776..114N}.}
    \label{fig:covering_frac}
\end{figure}
\begin{figure}
    \centering
     \includegraphics[width=\linewidth]{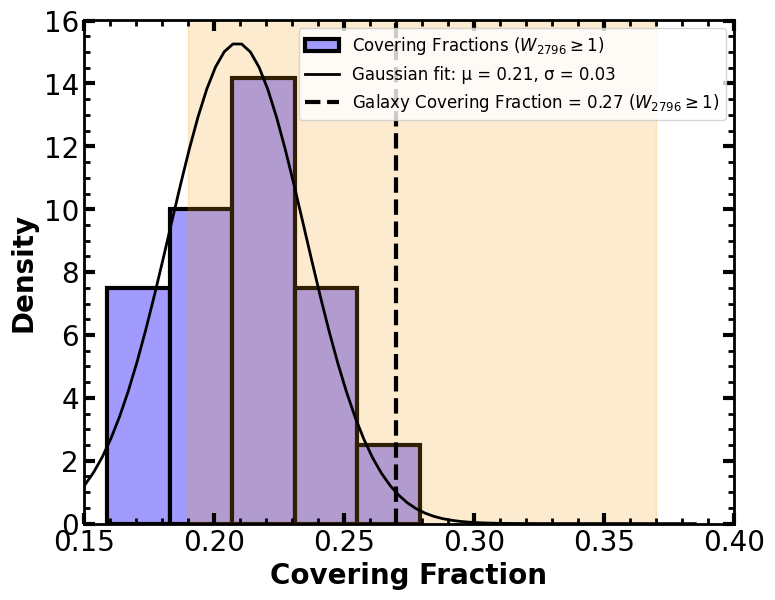}
    \caption{Comparison of the covering fraction distribution of quasars with galaxies, selecting quasars with impact parameters within a matching tolerance of 10 kpc for the galaxy sample, constrained by $D < 100$ kpc and foreground galaxies with redshifts in the range $0.4 \leq z \leq 0.8$. The vertical dashed line shows the covering fraction and the associated error for the galaxy sample with the same range of redshift and impact parameter.}
    \label{fig:covering_fraction_distribution}
\end{figure}


The covering fraction ($f_c$) is defined as the ratio of number of detections above a given rest equivalent width ($W_{2796}$) threshold to the total number of sightlines with sufficient sensitivity to detect \MgII\ absorption at that threshold. Accordingly, our $f_c$ estimates at a chosen $W_{2796}$ threshold include only those sightlines—whether detections or non-detections—that have a signal-to-noise ratio (SNR) adequate to detect the desired equivalent width at the 3$\sigma$ level.
In Figure~\ref{fig:covering_frac}, we present the covering fractions of \mgii\ absorbers for three different threshold of $W_{2796}$ in two impact parameter bins having equal number of sightlines. First, we observe that our sample follows the well-established trend of decreasing covering fraction  with increasing impact parameter and rest equivalent width threshold. 
For comparison we show the results obtained for \Magiicat\ galaxies \cite{Nielsen2013ApJ...776..114N} and results obtained by \citet{Chen2023ApJS..265...46C} for their full sample of galaxies.  Our covering fraction measurements are consistent with the literature values. However,
to draw a firm conclusion it is necessary to restrict the galaxy sample to 
the same redshift and impact parameter range as the quasar sample.

%
%
To achieve this, as before, we resample the quasar data to match the impact parameter distribution of the galaxy sample. We then calculate the average covering fraction within $D = 100~ kpc$. This process is repeated multiple times to generate a distribution of covering fractions.
The resulting covering fraction distribution for quasars is presented in Figure \ref{fig:covering_fraction_distribution}, along with the covering fraction for normal galaxies, represented by a vertical black dashed line. The shaded region around this line indicates the corresponding uncertainty. As evident from the figure, the covering fraction for quasar host galaxies remains well within the $1\sigma$ uncertainty of the covering fraction observed for normal galaxies.


\begin{figure}
    \centering
     \includegraphics[width=\linewidth]{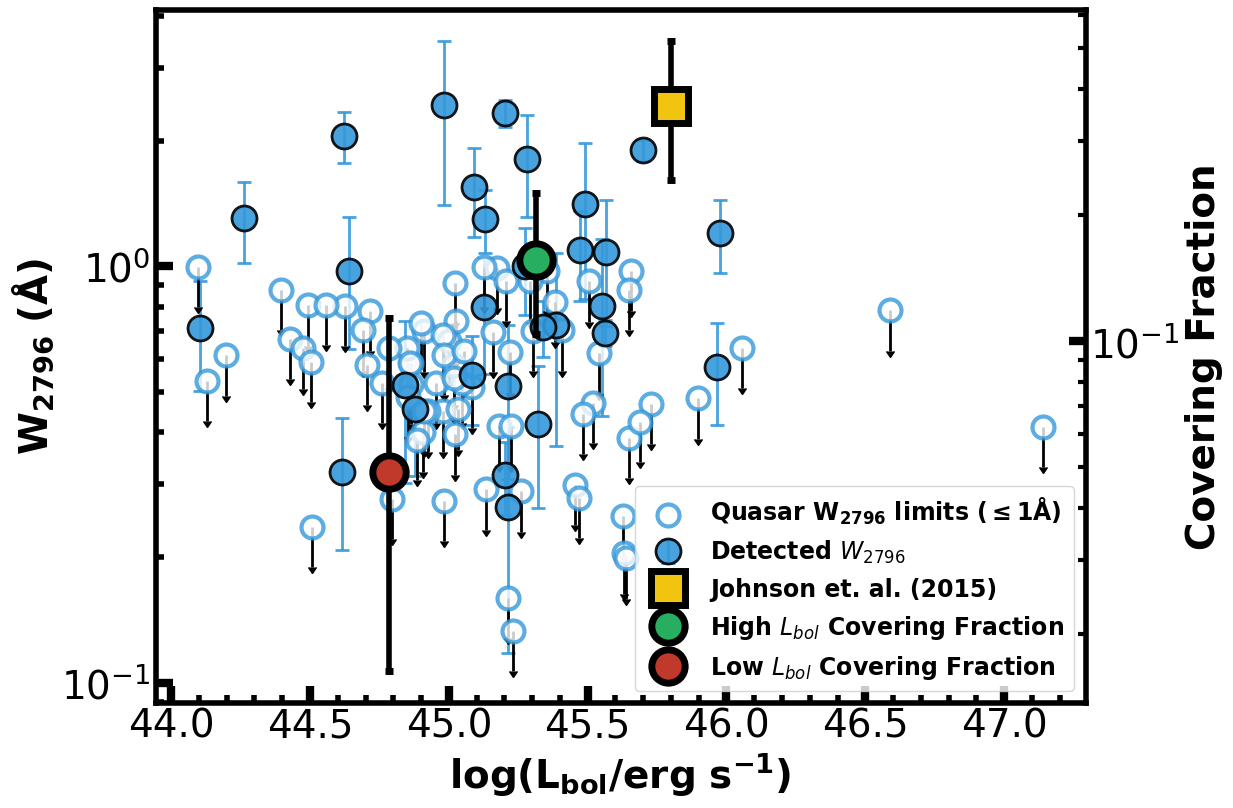}
    \caption{The figure illustrates the variation of \( W_{2796} \) with bolometric luminosity for the quasar sample. Overlaid in the right hand side ordinate are the covering fractions for two luminosity bins and $W_{2796} \ge 1$.}
    \label{fig:covering_frac_with_log_lbol}
\end{figure}

\citet{Johnson2015MNRAS.452.2553J} found a strong correlation between the covering fraction and bolometric luminosity. They also observed that the covering fraction measured at a given impact parameter is higher for quasar host galaxies with high luminosities (\(\log L_{\text{bol}}/erg~s^{-1} > 45.5\)) compared to normal galaxies. 
To explore this, in Figure~\ref{fig:covering_frac_with_log_lbol}, we plot \(W_{2796}\) as a function of \(L_{\text{bol}}\) where we have also included $3 \sigma$ limits for $W_{2796} \ge 1$\AA.
To assess the relationship between \(W_{2796}\) (including limits) and \(L_{\text{bol}}\), we applied a Gehan-Wilcoxon-type log-rank test resulting in test statistic of $0.81$ and $p_{null}$ of \(0.67\), indicating no significant dependence between \(W_{2796}\) and \(L_{\text{bol}}\).
Next, we divide the sample into two luminosity bins at \(\log L_{\text{bol}}/erg~s^{-1}=45\). The estimated covering fractions for absorbers with \(W_{2796} \geq 1\)~\AA\ are also shown in Figure~\ref{fig:covering_frac_with_log_lbol}. We observe a clear increasing trend in covering fraction with increasing \(L_{\text{bol}}\). Since we consider a restricted range of impact parameters and redshifts, the redshift distributions of the two luminosity bins are consistent with each other, as confirmed by a KS test yielding a $p_{null}$ of \(0.45\). Similarly, the KS test for the impact parameter distributions results in a $p_{null}$ of \(0.33\), indicating no significant difference between the two luminosity samples which can create a bias in the calculation of the covering fraction. Thus, we confirm a weak dependence of the covering fraction on bolometric luminosity. However, we do not find the measured covering fraction for any of the two bins being statistically different from that measured for the galaxy sample.

In Figure~\ref{fig:covering_frac_with_log_lbol}, we also include the covering fraction measured for the sample of \citet{Johnson2015MNRAS.452.2553J} for \(D < 100\) but without any restriction on the redshift range. Their measured average luminosity (see Figure~\ref{fig:luminosity_comparison}) and covering fraction (i.e., $0.36^{+0.15}_{-0.12}$) are higher than those obtained for our sample ($0.11^{+0.04}_{-0.03}$).
This once again indicates the increase in covering fraction with increasing bolometric luminosity. Note, the sample of \citet{Johnson2015MNRAS.452.2553J}
covers a wider redshift range and any dependence of $f_c$ on $z$ could affect our interpretation of this result. Also the covering fraction for the galaxy population (i.e $0.27^{+0.10}_{-0.08}$) is not significantly different than what is obtained for the sample of \citet{Johnson2015MNRAS.452.2553J}. {\it Thus our study indicates the need for a detailed investigation of these issues with larger samples well matched in luminosity, impact parameter and redshift distribution between normal galaxies and quasar host galaxies.}
%

\subsection{Influence of stellar mass} 

\begin{figure}
    \centering  \includegraphics[width=1.05\linewidth]{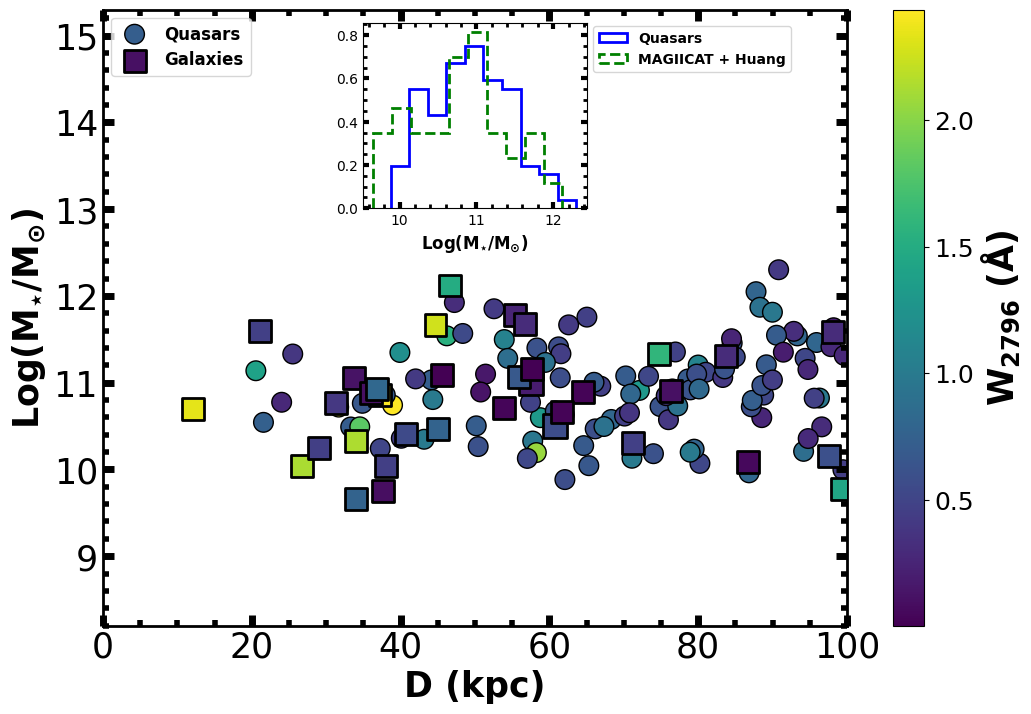}
    \caption{
    The plot illustrates the variation of stellar mass with the impact parameter, color-coded by the equivalent width of the detected \mgii absorption line. For quasar host galaxies the stellar mass is calculated using the relation given by \citet{decarali2010MNRAS.402.2453D}, as given in equation \ref{eq:stellarmass_blackholemass}.}
    \label{fig:stellar_mass}
\end{figure}

It is well documented in the literature that $W_{2796}$ vs. D as well as $f_c$ vs. D relationships depend on the stellar mass of the galaxy \citep[see for example,][]{Lan2020ApJ...897...97L}. So it is important to check how the stellar mass distribution of quasar host galaxies and normal galaxies used in this work compare with each other.
In Figure \ref{fig:stellar_mass}, we plot stellar mass vs. impact parameter for both the sample. This 
figure also compares the histogram distribution of stellar masses for both the samples.
For the galaxy sample, we calculated galaxy stellar masses only for cases where photometry was available in SDSS DR17 \citep{sdssdr172022ApJS..259...35A} or DECaLS \citep{decarali2010MNRAS.402.2453D} using the method described in \citet{Guha2022}.
For quasars, the stellar mass was estimated,
 using the empirical relation between black hole mass and host galaxy stellar mass from \cite{decarali2010MNRAS.402.2453D}, based on 96 quasars over redshifts 0.07–2.74, as:
\begin{equation}
\log\frac{M_{\mathrm{BH}}}{M_{\mathrm{host}}} 
= \bigl(0.28 \pm 0.06\bigr)\, z \;-\; \bigl(2.91 \pm 0.06\bigr).
\label{eq:stellarmass_blackholemass}
\end{equation} 
Here,  \( M_{host} \) represents the stellar mass, while \( M_{BH} \) denotes the black hole mass 
For SDSS quasars, \( M_{BH} \) values were obtained from the catalog of \cite{Wu2022ApJS..263...42W}. For quasars from the DESI EDR, we applied the same methodology as described in their work to derive \( M_{BH} \), ensuring consistency in stellar mass estimation.
As can be seen from the Figure \ref{fig:stellar_mass} that the stellar mass distributions are nearly same.
The median stellar mass for the quasar sample is found to be $10^{10.89} M_\odot$, whereas for the galaxy sample, it is $10^{10.86} M_\odot$. To statistically compare these distributions, we performed KS test, which resulted in a KS statistic of $0.170$ and a $p_{null}$ of $0.341$. Given the high $p_{null}$, we conclude that there is no significant difference between the two distributions. 
This result indicates that quasars in our sample reside in host galaxies that are not significantly more massive than the general galaxy population. Therefore, we can conclude that the similarity of the dependence of $f_c$ and $W_{2796}$ on D for both the samples is not influenced by a possible difference in the stellar mass distribution. 

\subsection{Line of sight velocity difference}
\label{sec:Line of sight velocity difference}
\begin{figure}
    \centering
    \includegraphics[width=\linewidth]{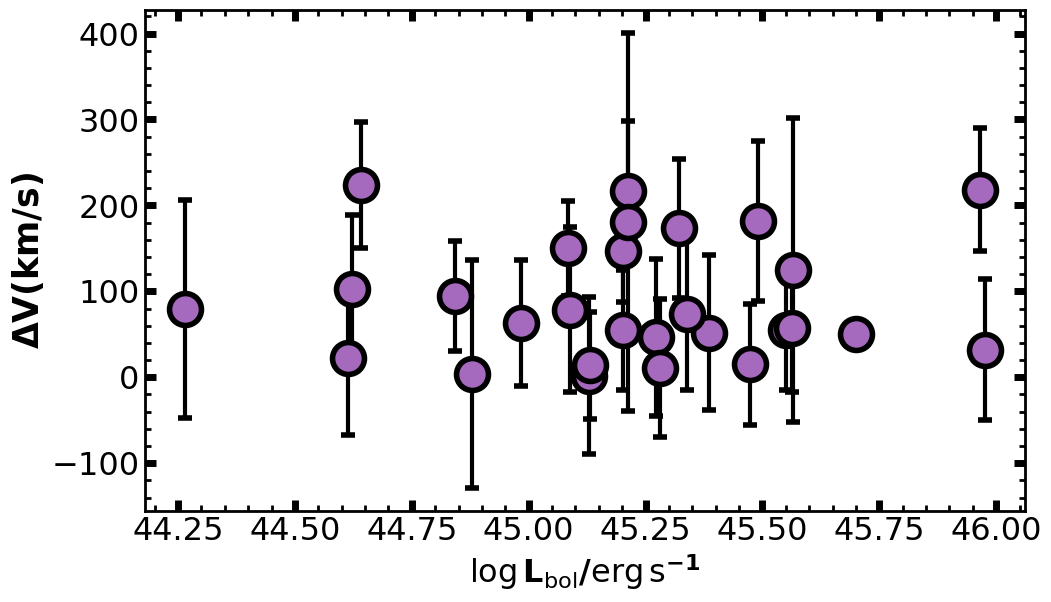}
    \caption{The figure presents the absolute velocity difference between the foreground quasar redshift and the absorber redshift as a function of the bolometric luminosity of the foreground quasar.}
    \label{fig:dvvsloglbol}
\end{figure}

\citet{Johnson2015MNRAS.452.2553J} indicated that the velocity off-set between the systemic redshift of the quasar and \MgII\ absorption redshift
have large spread compared to what has been found between galaxies and \MgII\ absorption.  We have already noticed that in our redshift and stellar mass matched samples apart from a small off-set the distribution is not 
statistically distinguishable between galaxies and quasar host galaxies.
Here we explore a possible correlation between the bolometric luminosity and line of sight velocity difference in our quasar sample.
In Figure \ref{fig:dvvsloglbol}, we present the velocity offset between the foreground quasar redshift and its bolometric luminosity. Our analysis reveals no significant correlation between these two quantities (Spearman Rank Correlation: $-0.009$, $p_{null}: 0.963$). Thus, within our sample, we find no clear evidence that the gas kinematics along the line of sight is influenced by the bolometric luminosity of quasars. The apparent discrepancy with \citet{Johnson2015MNRAS.452.2553J} may stem from their inclusion of absorption complexes—6 out of 73 total absorbers exhibit average velocity offsets of 721~km~s$^{-1}$—where multiple \mgii\ components are treated as distinct systems. This treatment contributes to the larger scatter in velocity offsets reported in their study. In contrast, our sample contains no such complexes, and we focus exclusively on strong \mgii\ systems, which are more likely to originate from regions closer to the galactic center.

\section{Discussion and Conclusion}

It is now well established that the circumgalactic medium (CGM) consisting of metal enriched diffuse gas is a crucial component of a galaxy.
The CGM serves as a reservoir of gas that regulates star formation and accretion processes. The properties of CGM are probed either through absorption lines detected in the spectra of background quasars or through the detection of diffuse emission.  
What is the influence of the central AGN on the properties of CGM? This is an important question to address \citep[see ][for using \MgII\ absorption as a tool]{Farina2014MNRAS.441..886F, Johnson2015MNRAS.452.2553J,Chen2023ApJS..265...46C}.  These studies have indicated that rest equivalent width of \MgII\ absorption and its detection rate (or covering fraction) tend to be higher around AGN host galaxies compared to the normal galaxies. It is also well documented that the correlation between $W_{2796}$ and impact parameter has a large scatter and may depend on host  galaxy properties such as (i) stellar mass, (ii) luminosity  and (iii) redshift etc. Therefore, it is important to probe the influence of quasars using samples that are well matched in various properties of galaxies and impact parameter distribution. This forms  the main motivation of this work.
%

Here, we have utilized a sample of well selected 166 closely spaced quasar pairs, to probe the CGM of quasar hosting galaxies in  comparison to the sample of the CGM of normal galaxies in redshift range of 0.4\(~\le z\le~\)0.8 and impact parameter range of 20-100 kpc (section 2). Based on the typical SNR achieved in the SDSS spectra, we mainly focus on systems having  \(W_{2796} \ge 1\)\AA. As described in Section \ref{sec:sample_selection}, our comparison galaxy sample is an amalgamation of  MAGICAT~\citep{Nielsen2013ApJ...776..114N} and \cite{Huang2021MNRAS.502.4743H} sample, matching both in redshift as well as in \(D\).
Our main result is that both  \(W_{2796}\) versus impact parameter (\(D\)) distribution and covering fraction (\(f_c\)) versus  \(D\) distribution are statistically similar 
among the normal galaxies and  quasar host galaxies in the aforementioned redshift and impact parameter ranges. We ensured that our results are not biased by the different impact parameter distribution of the two samples. In addition, the stellar mass distributions used for the galaxy sample and quasar host sample are statically similar (e.g., Figure~\ref{fig:stellar_mass}), precluding any bias due to stellar mass dependence. A recent study by \citet{Shibata2025arXiv250504259S} found that the mean local galaxy density around quasars is $\sim11$--$20\%$ lower than that around matched galaxies on scales of $0.3$--$0.7$\,pMpc. Although their analysis probes larger scales, this suggests that clustering does not significantly influence the absorbers in our study, supporting the interpretation that our results are independent of large-scale environmental effects.
\par

Spliting our sample into two bolometric luminosity bins we do see a weak dependence of $f_c$ on luminosity, however no correlation was present between the bolometric luminosity and $W_{2796}$. This appears to be inconsistent with the result of
\citet{Johnson2015MNRAS.452.2553J}, who have reported enhanced   \(W_{2796}\) and $f_c$ for luminous quasar host galaxies compared to low-luminosity quasars and inactive galaxies. However, it may be noted that quasars in our sample typically probe the low bolometric luminosity end of the quasar sample studied by \citet{Johnson2015MNRAS.452.2553J} and  their sample probes a much wider redshift and impact parameter ranges which make direct comparison inappropriate.
Moreover to confirm the strong trend in $f_c$ with $L_{bol}$
we need to include sample of high luminosity quasars over the same redshift range to our sample. This will alleviate the redshift evolution effects influencing the result. 
It is also important to extend our analysis to lower rest equivalent width limits to see whether the discrepancy is related to us mainly focusing on $W_{2796}\ge1$\AA\ absorbers. This can be done by freshly obtaining higher SNR spectra of quasars in our sample.

Similarly, the study by \citet{Chen2023ApJS..265...46C}, used large sample over full redshift range probed by SDSS to stack the background spectrum at foreground quasar redshift and showed that the  \(W_{2796}\)  is relatively higher in comparison to \(W_{2796}\) from CGM of normal galaxies. It will be important to revisit their analysis with a redshift and impact parameter matched samples of quasar host galaxies and normal galaxies in a restricted redshift range. 
\citet{Chen2023ApJS..265...46C} have also noticed that detection probability of \MgII\ absorption along the line of sight is less than that along the transverse direction. 
We confirm this result in our analysis as well. Interestingly, in the case of normal galaxies  the values of \(W_{2796}\) as well as \(f_c\) is found to increase with decreasing impact parameter \citep[see for example,][]{Guha2024}. Therefore we expect both the quantities to be higher along the foreground quasar line of sight (probing also at low impact parameters) compared to what we find in the transverse direction (probing at higher impact parameters).  In addition, one expects contribution of associated \MgII\ absorption also to be present along the line of sight. Therefore, 
less detection of \MgII\ absorption along the line of sight may be related to anisotropy either in the radiation field or matter distribution in the case of quasar host galaxies \citep[see for example,][]{Jalan2019ApJ...884..151J}. Therefore, our finding of similarity in the CGM properties over the impact parameter range 20-100 kpc is intriguing. As mentioned before, we do not have many lines of sight with impact parameter $<50$ kpc. It will be important to construct such a sample to confirm the presence of anisotropy in the case of quasars.


\citet{Johnson2015MNRAS.452.2553J} have also reported an increase in the velocity off-set between the quasar host galaxy and absorption redshift with bolometric luminosity. This was also considered as one of the indications of influence of the quasar on the CGM. However, we
 do not find the scatter in the velocity offset values to increase with increasing luminosity of quasar host galaxies (e.g., see Figure \ref{fig:covering_frac_with_log_lbol}). 
One possible explanation for the larger velocity offsets reported in \citet{Johnson2015MNRAS.452.2553J} could be the presence of multiple absorption complexes within their sample. However, this appears unlikely, as only 6 out of the 73 absorbers are associated with such complexes, and their exclusion reduce r.m.s  of the velocity offset only from 496 km/s to 465 km/s.
It is therefore plausible that other factors, such as larger impact parameters, higher quasar host galaxy luminosities, and higher redshift range, may partially account for the discrepancy.

To conclude, our detailed analysis based on a sample of 166 quasars in comparison with a  redshift and impact parameter matched control sample of normal galaxies,  show no significant difference in the \(W_{2796}\)–\(D\) and \(f_c\)–\(D\) distributions among them, indicating that low-luminosity quasars (i.e., mostly \(L_{bol}\le 10^{45} \) erg s$^{-1}$) do not substantially alter the CGM properties of the host galaxies. 
For further progress, future studies should focus on observing sight-lines selected in a ``galaxy-first'' manner, using large samples to enable more accurate comparisons with quasar host galaxies. In parallel, CGM studies of quasar hosts can be improved by targeting: (i) luminous quasars, (ii) high signal-to-noise ratio (SNR) data to detect low \( W_{2796} \) values, (iii) high spectral resolution to probe detailed kinematics and (iv) a range of impact parameters (including \(D < 50\) kpc as well) by carrying out controlled analysis similar to our study here. 
\label{sec:discussion}

\section{Acknowledgements}
We thank the anonymous referee for constructive comments and suggestions, which helped improve the clarity and quality of this manuscript.
The research of P.S. is supported by the University Grants Commission (UGC), Government of India, under the UGC-JRF scheme  (Ref. No.: 221610014755). H.C. and P.S. express their gratitude to the Inter-University Centre for Astronomy and Astrophysics (IUCAA) for their hospitality and the provision of High-Performance Computing (HPC) facilities under the IUCAA Associate Programme. We also thank Sowgat Muzahid and Sayak Dutta for their valuable discussions during this project. Additionally, we acknowledge the assistance of AI tools, specifically OpenAI's ChatGPT, for aiding in writing and code development, and Grammarly for enhancing the text's clarity and correctness.

\section*{Data Availability}
The data used in this study are publicly available in the DESI-EDR, and SDSS DR16 Data Release.



\bibliographystyle{mnras}
\bibliography{references} 




\appendix


\bsp	
\label{lastpage}
\end{document}